\documentstyle[prb,aps,epsf]{revtex}
\begin{document}
\draft
\title{The instrument response function in air-based scanning tunneling microscopy}
\author{P. Fraundorf and J. Tentschert}
\address{Physics \& Astronomy, U. Missouri-StL (63121), \\
Corporate Research, Monsanto (63167), \\
Physics, Washington U. (63130), \\
St. Louis, MO, USA}
\date{\today }
\maketitle

\begin{abstract}
The distinction between point and line resolution in transmission electron
microscopy (TEM) arises because an ability to image sub-0.2 nm fringes is a
necessary, but not a sufficient, condition for imaging individual atoms. In
scanned tip microscopy, as in TEM, empirical data on instrument response
should precede assertions about point resolution. In the ``slow scan
limit'', time-domain noise and geometry effects decouple, and tip shape can
take on the role of a 2-dimensional impulse response function. We indicate
here that nuclear track pits can be used to quantitatively measure tip
geometry with nanometer-scale resolution {\em in three dimensions}, that
stationary tip images provide a {\em robust} measure of time-domain
instabilities, and that when these data are taken before and after imaging
an unknown, images with instrument response quantitatively constrained by
experiment are possible. Specimen-induced tip effects also become measurable 
{\em in situ}. Also {\em Ultramicroscopy} {\bf 37} (1991) 125-129.
\end{abstract}




\section{Introduction}

In the literature of air-based scanned tip microscopy, images with $0.2$ nm
(or larger) fringes are often described as ``atomically-resolved''.\cite
{Colton} The assumption that fringe spacings provide an upper limit on point
resolution in tunneling images was plausible, prior to the availability of
recent evidence: (a) that tunneling in air or poor vacuum generally involves
mechanical contact between specimen and tip,\cite{Spence1} (b) that
``corrugation depths'' between $0.2$ nm rows in tunneling images of $\langle
001\rangle $ graphite and other layered structures can be many tenths of nm,%
\cite{Hansma} (c) that the above images can become increasingly noise-free
as tip sharpness decreases and mechanical contact between specimen and tip
increases, and (d) that scanning force images unlikely to have point
resolutions on the atomic scale also show atomic scale periodicities.\cite
{Bryant} Mechanical contact between specimen and tip (always present in
repulsive van der Waals force microscopy) leaves open the possibility that a
periodic raft of specimen atoms, and not the tip itself, is sliding across
the specimen, averaging periodicity information over distances much larger
than a single unit cell. The ease of imaging fringes on layered structures
thus provides evidence that lattice fringes alone, as in transmission
electron microscopy, provide no proof of point resolution alone.

For this reason, and because the instrument response function in scanned
probe microscopy introduces artifacts in the time domain as well as in
space, researchers involved in air-based scanned tip studies should be at
least as cautious as electron microscopists\cite{Spence2} in documenting
their instrument's response. Because of the serial mode of data acquisition
in scanning tunneling, spatial and temporal aspects of instrument response
may require separate measurements. We show here how the elements for
characterizing time domain instrument response are already present in the
literature, and how nuclear particle track pits (etched\cite{FPW} or unetched%
\cite{Fraundorf1}) provide a means for measuring the three-dimensional
interaction profile for a given tip, while on site in the microscope.

\section{Components of instrument response}

Measuring instruments generally start with some property associated with a
specimen, and end up producing a data object for analysis. For example, the
contrast transfer function in high-resolution TEM begins with the electron
wave-function at the exit surface of the specimen, and produces an image
which may contain partial information on exit surface deBroglie phase from
point to point. In scanned probe instruments, one begins with the profile
from point to point of another quantity (e.g. the piezo height $z(x,y)$
associated with current $I$ at bias voltage $V$). From the resulting image
we seek information on the {\em local} profile of some height $h$ (i.e., of
constant current for an arbitrarily sharp tip at constant conductivity $I/V$%
). We refer here to instrument response as that function $R$ which maps the
local profile $h(x^{\prime },y^{\prime })$ into the collected image $z(x,y)$%
, i.e., it is the function for which $z(x,y)=R(x,y,\{h(x^{\prime },y^{\prime
})\forall x^{\prime },y^{\prime }\}$).

Because the $z(x,y)$ values are obtained sequentially in time, the geometric
aspects of $R$ are intertwined with time-domain components of instrument
response which are at best partially known. However, in the slow scan limit,
for which the tip is allowed to assume a steady state with each new value of 
$x$ and $y$ (i.e., when the feedback loop is fast enough to follow specimen
topography for the scan rates in use), the geometric and time domain effects
can be separated, and $z(x,y)$ can be written as:

\begin{equation}
z(x,y)=R_g\left( x,y,\left\{ h(x^{\prime },y^{\prime })\forall x^{\prime
},y^{\prime }\right\} \right) +N(x,y)\text{,}  \label{IRF}
\end{equation}
where $R_g$ is a strictly geometric version of the instrument response, and $%
N(x,y)$ respresents the time-domain noise in the system. The next section
discusses procedures for measuring the second moment statistics of $N(x,y)$,
as well as Bayesian (i.e., best guess) strategies for subtracting $N(x,y)$
from images therewith. The remaining function $R_g$ depends only on the
details of tip structure, and of course on the signal being measured. In the
simplest case of a profile which is: (i) dominated by tunneling current
through the infinitesimal point on the tip nearest the specimen, and (ii)
dependent on gap distance but not on local curvature, composition, etc., $%
R_g $ takes the form\cite{Moller} of

\begin{equation}
R_g(x,y)=h(0,0)+\max \left[ \left\{ h(x+x^{\prime },y+y^{\prime
})-t(x^{\prime },y^{\prime })\right\} \forall x^{\prime },y^{\prime }\right] 
\text{.}  \label{Simplest}
\end{equation}
Here $t(x^{\prime },y^{\prime })$ are $z$-coordinates of the tip surface
while it is pointing toward a specimen in the negative $z$-direction, with
the tip positioned horizontally so that its scan coordinates are $\left\{
x,y\right\} =\left\{ 0,0\right\} $, and vertically so that $t(0,0)$ equals
the height of the specimen at the scan origin. In practice, of course, we
ignore absolute heights and consider relative heights only. Note here that $%
t(x^{\prime },y^{\prime })$ serves as a kind of impulse response function,
since it predicts the shape of the noise-subtracted image for the limiting
case of an infinitely sharp specimen. In low-magnification air-based
scanning tunneling microscopy on rough specimens, where tip geometry has
major effects and the nature of tip-specimen interaction is poorly known,
measured profiles of the effective tip-interaction geometry $%
t_{eff}(x^{\prime },y^{\prime })$ provide useful estimates of the impulse
response in these images as well. A simple method for measuring profiles of
tip-interaction geometry in the microscope, on the nanometer scale, is
therefore discussed in the section on tip fingerprints below.

\section{Stationary tip images}

Before discussing use of stationary tip images to determine the power
spectrum of N(x,y), we should first discuss violations of the slow scan
assumption. Time-domain artifacts associated with response of the feedback
loop to slope changes on the specimen include spikes, associated with rapid
encounters by the tip with slopes that rise too quickly, and ``hand-glide''
artifacts associated with slopes which drop away too soon. Imaging of
objects more than once, with different scan rates and amplitudes, can help
detect features like these, and comcomitant violation of the assumptions
which go into eq. (\ref{IRF}). Slowing the scan rate, of course, entails
compromise itself, not the least of which in air is the amount of thermal
drift that takes place during image acquisition.

The statistical properties of $N(x,y)$ in eq. (\ref{IRF}), given its
appropriateness for a specific set of images, have been examined, and
statistical removal by one-dimensional\cite{Park} and two-dimensional\cite
{Stoll} Wiener optimal filtering has been proposed provided a suitable model
for the noise power spectrum is available. These filtering techniques are a
special case of Bayesian background subtraction, a technique which is now
finding application in place of Fourier averaging for the removal of noise
from HREM images as well.\cite{Fraundorf2} For the empirical background
model in STM, Stoll and Marti\cite{Marti} have proposed using {\em %
stationary tip images} (defined here as images taken with the $x$ and $y$
tip scan-range set to zero). These stationary tip images are of special
importance because, like HREM images of aperiodic material used in
quantifying instrument contrast transfer, they can be obtained: (i) on the
same specimen, (ii) under the same imaging conditions, and (iii) at {\em %
almost} the same time as the image whose instrument response is being
considered.

We note in passing that stationary tip images, also like HREM images of
aperiodic material, provide a robust measure of the effect of time-domain
instability. They are quite sensitive to changes in instrument and
environment, but insensitive to changes in tip and often feedback loop
settings.\cite{Fei} Hence they can be used to measure instrument performance
as well as the effects of site instability on that performance.

\section{Fingerprints of the tip}

After removal of time-domain artifacts with help from stationary tip and
multiple-scan images, the task of relating $z(x,y)$ to some local property
of the specimen remains. In HREM, the analogous setp involves determining
the contrast transfer function for a given image, with help from aperiodic
structure in the image as well as prior information on instrument response.
A specimen-independent fingerprint of instrument response in scanned probe
images does not appear to be available. The empirical approach is therefore
simple: constrain the tip response before and after taking an unknown image
by observations on a known.

The simplest {\em known} for this purpose would be an otherwise flat surface
populated with {\em identical} Poisson-distributed delta-functions (i.e.,
atomically sharp peaks). In principle such specimens can exist for both
tunneling and force microscopy, but in practice they remain to be
identified. However, the nuclear particle track formation mechanism can
already provide geometrically uniform, nanometer-sized damage regions
capable of systematic enlargement by etching to diameters well over a micron.

Fig. 1 shows an image obtained by averaging scanning tunneling images of
four nearby $50$ nm diameter etched and Au-Pd coated nuclear particle track
pits.\cite{Tentschert} Individual data points in this pit profile showed
standard deviations of only $2$ nm, and the standard deviation {\em image}
indicated that spatial correlations in these errors, if any, were confined
to regions of steep slope. The contours in Fig. 1 represent levels of {\em %
three standard deviations} in the average rms pixel spread from image to
image. A well defined and repeatable structure of ledges is present.
Resolution is near to (or better than) $5$ nm in all directions, and we
suspect this is limited in part by grain size in the Au-Pd coating, and in
part by variations (obviously quite small) from pit to pit.

This geometric pit profile of Fig. 1 can be inverse-transformed (see inset)
to provide a three-dimensional view of those portions of the tip-interaction
profile which we can geometrically (and/or electronically) bring into
contact with the inside of a $50$ nm diameter ring.\cite{Fraundorf3} In
other words, quantified pit images like those shown in Fig. 1 provide
``better-than-convex envelope' characterizations of $t_{eff}(x^{\prime
},y^{\prime })$, in this case with {\em quantitative} spatial resolution 
{\em in three dimensions} on the order of one-tenth of the hole diameter.
These pit images can be obtained before and after imaging an unknown
specimen, thus allowing one to characterize tip modifications as well as tip
structure.\cite{Tentschert2} Use of smaller and larger pits allows
characterization of tip geometry on smaller and larger size scales as well.
We expect the strategy will allow formation of reproducible pits in the nm
size range for scanning force work, although as pit size decreases for
scanning tunneling work the problem of texture in the conductive coating is
expected to increase in severity.

\section{Conclusions}

It is, of course, possible to compare unknown specimen images with other
specimen images, and with model expectations based on prior information, and
thereby to do a great deal of microscopy without hard information on
instrument response. In vacuum-based scanned probe microscopy, prior
information about both specimen and tip may be adequate to obviate the need
for empirical data on instrument response. In air- and liquid- based scanned
probe study, however, a strength of the technique is that the instruments
are capable of examining true unknowns. But before structure images, in the
electron microscopic sense\cite{Spence2}, can be obtained in general, an
empirical protocol for inferring (and removing) the effects of instrument
response is necessary.

We have illustrated how theoretically, in the slow-scan limit, the
time-domain and geometric components of scanned probe instrument response
can be decoupled, and how the geometric interaction profile of the scanned
tip in that case can become a two-dimensional impulse response function.
Further, we have shown how stationary tip images may serve as a robust
source of information on time-domain instabilities, obtainable while the tip
is in place on the unknown of interest. Finally, we illustrated that the
geometric profile of a tip in air (or liquid or vacuum) can be measured with 
{\em 5} nm or better resolution, {\em in the microscope} before and after
examination of an unknown specimen. This confirms our earlier observation
that nuclear particle tracks are a promising tool in this regard, and
further opens the door to quantitative studies of the effect of specimens on
tips: a needed complement when tip-induced changes to the specimen are being
considered. Overall, feeling the need because we have been ``spoiled'' by
elegant tools for characterizing instrument response in electron phase
contrast imaging, we: (i) have proposed a strategy for obtaining data on
scanned probe instrument response {\em almost} on an image-by-image basis,
and (ii) where we have not seen the components used elsewhere have
illustrated their workability.

\begin{figure}[tbp]
\caption{Image obtained by averaging $32\times 32$ pixel images of four $50$
nm diameter etched, and Au-Pd coated, nuclear particle track pits in
polycarbonate. The pit images were cut from larger $12$-bit images obtained
with an $8$ $\mu $m Digital Instruments head using a single mechanically
formed Pt-Ir tip. The bottom of the pit is white, while the surrounding flat
polycarbonate is dark. Contours (at intervals of 3 standard deviations in
the rms pixel variation from image to image) are white in regions
surrounding the pit, and dark in the deeper regions of the structure. The
pit base contour (white) is round and approximately 50 nm in diameter,
although it appears oval here because of different magnifications in the x
and y scan directions. Depth from base to bottom is of the order of 20 nm
(around 487 greyvalue units), based on rough z-gain calibrations using 30 nm
collodial gold on graphite. The inset shows the resulting envelope (z-height
exaggerated) for the 20 nm end of the tip used to take this data (see text).}
\label{Fig1}
\end{figure}

\end{document}